\documentclass{article}
\usepackage{spconf,amsmath,graphicx}
\usepackage{cite}
\usepackage[linesnumbered, ruled]{algorithm2e}

\let\oldnl\nl
\newcommand{\nonl}{\renewcommand{\nl}{\let\nl\oldnl}}

\title{JOINT ML CALIBRATION AND DOA ESTIMATION WITH SEPARATED ARRAYS}
\name{V. Ollier$^{\star}$ $^{\diamond}$ \qquad M. N. El Korso$^{\dagger}$ \qquad
R. Boyer$^{\diamond}$ \qquad P. Larzabal$^{\star}$ \qquad M.
Pesavento$^{\ddag}$\thanks{This work was supported by the
following projects: MAGELLAN (ANR-14-CE23-0004-01), MI-CNRS TITAN
and ICode blanc.}}
\address{$^{\star}$ SATIE, UMR 8029, ENS Cachan, Universit\'{e} Paris-Saclay, France \\
$^{\dagger}$ LEME, EA 4416, Universit\'{e} Paris-Ouest, Ville d'Avray, France \\
$^{\diamond}$ L2S, UMR 8506, Universit\'{e} Paris-Sud,
Gif-sur-Yvette, France \\ $\ddag$ Communication Systems Group,
Technische Universit\"{a}t Darmstadt, Darmstadt, Germany}
\begin{document}
\ninept
\maketitle
\begin{abstract}
This paper investigates parametric direction-of-arrival (DOA)
estimation in a particular context: i) each sensor is
characterized by an unknown complex gain and ii) the array
consists of a collection of subarrays which are substantially
separated from each other leading to a structured noise covariance
matrix. We propose two iterative algorithms based on the maximum
likelihood (ML) estimation method adapted to the context of joint
array calibration and DOA estimation. Numerical simulations reveal
that the two proposed schemes, the iterative ML (IML) and the
modified iterative ML (MIML) algorithms for joint array
calibration and DOA estimation, outperform the state of the art
methods and the MIML algorithm reaches the Cram\'{e}r-Rao bound
for a low number of iterations.

\end{abstract}
\section{Introduction}
\label{sec:intro}

%%%%hawkes1998acoustic,
Direction-of-arrival (DOA) estimation \cite{haardt2014subspace,
stoica1990maximum} is an important topic with a large number of
applications: radar, satellite, mobile communications, radio
astronomy, geophysics and underwater acoustics
\cite{godara1997application, wong1999root, van2013signal}. %It can be tackled with high
%resolution estimation techniques such as Weighted Subspace Fitting
%(WSF) \cite{viberg1991sensor}, MUtiple SIgnal Classification
%(MUSIC) \cite{stoica1989music} or Estimation of Signal Parameters
%via Rotational Invariance Techniques (ESPRIT) \cite{roy1989esprit}
%for example. %To benefit from a relatively low computational
%complexity, subspace based estimation techniques
%\cite{haardt2014subspace} or search-free techniques
%\cite{gershman2010one} can be well adapted.
In order to achieve high resolution, it is common to use arrays
with large aperture and/or a large number of sensors, in a
specific noise environment. Considering spatially and temporally
uncorrelated zero-mean Gaussian processes is a typical noise
assumption but it may be violated in numerous applications, as in
the context of sonar, where correlated or colored noise is
required \cite{ye1995maximum,
friedlander1995direction, stoica1996maximum, li2008maximum}.
%%%le1989parametric,
%
%This is notably considered for partly calibrated arrays
%\cite{pesavento2002direction, see2004direction,
%parvazi2011direction} with rank reduction estimators, as RARE. In
%this case, each subarray is assumed to be well calibrated, however
%due to uncertainties in the subarray displacements, the relative
%phase relation among subarrays is undetermined. Consequently, it
%is assumed that the array is partly and not fully uncalibrated.
%Furthermore, the methods in \cite{pesavento2002direction,
%see2004direction, parvazi2011direction} belong to the class of
%subspace-based DOA estimation methods which usually perform well
%for spatially and temporally uncorrelated zero-mean Gaussian
%processes. However, such noise assumption may be violated in
%numerous applications, as in sonar context, and require to
%consider correlated or colored noise \cite{le1989parametric,
%ye1995maximum, friedlander1995direction, li2008maximum}.

We consider here the case where the noise covariance matrix
exhibits a particular (block-diagonal) structure
\cite{vorobyov2005maximum} that differs from the classical
assumption: spatially white uniform noise
\cite{stoica1990performance, KorsoSeq} or non-uniform noise
\cite{pesavento2001maximum}. In our paper, we consider DOA
estimation in large sensor arrays composed of multiple subarrays.
Due to the large spacing between subarrays, we assume that the
noise among sensors of different subarrays is statistically
spatially independent. In a given subarray, however, the noise is
spatially correlated between sensors. This entails a block-diagonal
structure of the noise covariance matrix, linked to the sparsity
of the array.

Apart from this noise assumption, we also consider that in
realistic scenarios, due to miscalibration, the individual sensor
outputs are generally subject to distortions by constant
multiplicative complex factors (gains). These calibration errors
are hardware related in our case, leading to different DOA
independent sensor gains \cite{weiss1990eigenstructure,
vorobyov2005maximum}. To precisely estimate these errors, we take
advantage of the presence of calibration sources
\cite{pierre1991experimental, ng1996sensor} to simultaneously
calibrate and estimate DOAs. Our scenario is general and it can be
adapted or extended to some practical applications as in the radio
astronomy context \cite{wijnholds2010fish} where the constant
complex sensor gain assumption is common.
%Contrary to \cite{pesavento2002direction, see2004direction,
%parvazi2011direction}, we consider fully uncalibrated arrays and
%we take advantage of the presence of calibrator sources
%\cite{ng1996sensor, pierre1991experimental}. Our scenario is
%general but it could be adapted or extended to some practical
%applications as in the radio astronomy context
%\cite{wijnholds2010fish}.
% Indeed,
%the array we consider is linear and non uniform, but the extension
%to a planar array, like new interferometric radio telescopes as
%the low frequency array (LOFAR), is straightforward and is not
%presented here for sake of clarity.

We use the conditional/deterministic model
\cite{ottersten1993exact} for the signal sources. Nevertheless,
following the same methodology, we can adapt the proposed
algorithms to the case of unconditional/stochastic model
\cite{ottersten1993exact, stoica1995concentrated,
gershman2002stochastic, chen2008stochastic}. The two parametric
algorithms we present are based on the maximum likelihood (ML)
estimation method, due to its good statistical performances. The
size of the unknown parameter vector being large, we perform
iterative optimization to make the ML estimation problem
computationally tractable. Furthermore, the estimation
performances are improved with the introduction of calibration
sources in our scenario. To assess the performances
\cite{korsocond}, the
Cram\'{e}r-Rao bound (CRB) is used.%In our context, one needs to take into account the

The notation used through this paper is the following: scalars,
vectors and matrices are represented by italic lower-case,
boldface lower-case and boldface upper-case symbols, respectively.
The symbols $\left( \cdot \right) ^{T}$, $\left( \cdot \right)
^{\ast }$, $\left( \cdot \right) ^{H}$, $\left( \cdot \right)
^{\dag}$, $\mathrm{tr}\left\{ \cdot \right\} $ and $\det \left\{
\cdot \right\} $ denote, respectively, the transpose, the complex
conjugate, the hermitian, the pseudo-inverse, the trace and
determinant operator. The real and imaginary parts are referred to
by $\Re\left\{ \cdot \right\} $ and $\Im\left\{ \cdot \right\} $.
The operators $\mathrm{bdiag}\left\{ \cdot \right\}$ and
$\mathrm{diag}\left\{ \cdot \right\}$ represent a block-diagonal
and a diagonal matrix, respectively. A vector is by default a
column vector and $\mathbf{I}$ is the identity matrix.
%Finally, entries of a vector, respectively a matrix, are denoted
%by $\left[ \cdot \right] _{l}$, respectively $\left[ \cdot \right]
%_{h,l}$.
The symbol $\odot$ denotes the Schur-Hadamard product,
$\delta\left(.\right)$ is the Dirac's delta function and
$\mathbf{E}_{p}$ is a $p \times p$ matrix filled with ones.

\section{Observation model}
\label{sec:format}

We consider $D$ signal sources impinging on a linear (possibly not
uniform) array of $M$ sensors. The array response vector for each
source $l=1,\ldots ,D$ is defined as \cite{krim1996two}
% $\mathbf{a}\left( \theta_{l}
%\right) =[1,e^{\left(-j 2\pi f\frac{d_{2}}{c}\sin \left(
%\theta_{l}\right)\right)},\ldots, e^{\left(-j 2\pi
%f\frac{d_{m}}{c}\sin \left( \theta_{l}\right)\right)}]^{T}$
\begin{equation}
 \mathbf{a}(\theta_{l})=[1,e^{-j 2\pi f\frac{d_{2}}{c}\sin(\theta_{l})},\ldots, e^{-j 2\pi
f\frac{d_{M}}{c}\sin(\theta_{l})}]^{T}
\end{equation}
in which $\theta_{l}$ is the DOA of the $l^{th}$ source, $f$
denotes the carrier frequency, $c$ the propagation speed and
$d_{k}$ the inter-element spacing between the first and the
$k^{th}$ sensor. We note as $\lambda $ the wavelength of the
incident wave. The output observation of the full array is given
at each snapshot by
\begin{equation} \label{model1}\mathbf{y}(t)=\mathbf{A}(\boldsymbol{\theta }) \mathbf{s}(t)
+\mathbf{n}(t), \ \ t=1,\ldots ,N
\end{equation}
where $N$ is the total number of snapshots, $\boldsymbol\theta
=[\theta_{1}, \ldots, \theta_{D}]^{T}$ is the DOAs vector,
$\mathbf{s}(t)=[s_{1}(t),\ldots ,s _{D}( t)]^{T}$ the signal
source vector, $\mathbf{n}(t)=[n_{1}(t),\ldots ,n _{M}( t)]^{T}$
the additive noise vector and $\mathbf{A}(\boldsymbol{\theta }) =[
\mathbf{a}(\theta _{1}),\ldots ,\mathbf{a}(\theta _{D})]$ the
array response matrix. In matrix notation, we have
\begin{equation}
\label{model2} \mathbf{Y}=\mathbf{A}(\boldsymbol{\theta })
\mathbf{S}+\mathbf{N}
\end{equation}
with $\mathbf{Y}=[\mathbf{y}(1),\ldots ,\mathbf{y}(N)]$,
$\mathbf{S}=[\mathbf{s}(1),\ldots,\mathbf{s}(N)]$ and
$\mathbf{N}=[\mathbf{n}(1),\ldots ,\mathbf{n}(N)]$. In this work,
the different assumptions that we consider are the following:\\

%\subsection{Calibrator sources}
%\label{ssec:subhead1}
\textbf{\textit{A1) Calibration sources}}: In a number of
practical applications, the knowledge of one or multiple
calibration sources is available \cite{ng1996sensor,
boyer2008oblique, bouleux2009optimal, korso2009fast}. Without loss
of generality, we consider the first $P$ sources as calibration
sources with known DOAs. Thus, the steering matrix is partitioned
as
\begin{equation}
\label{partA} \mathbf{A}(\boldsymbol{\theta })=\big[\mathbf{
A}(\boldsymbol{\theta}_{K}),\mathbf{A}(\boldsymbol{\theta}_{U})\big]
\end{equation}
in which $\boldsymbol{\theta}_{K}=[\mathbf{\theta}
_{1},\ldots,\mathbf{\theta}_{P}]^{T}$ represents the known DOAs
and $\boldsymbol{\theta}_{U}=[\mathbf{\theta}_{P+1},\ldots
,\mathbf{\theta}_{D}]^{T}$ is the vector of the unknown DOAs.
Likewise, the signal source matrix can be written as follows
\begin{equation}
\label{partS}
\mathbf{S}=\big[\mathbf{S}_{K}^{T},\mathbf{S}_{U}^{T}\big]^{T}
\end{equation}
in which $\mathbf{S}_{K}=[\mathbf{s}_{K}(1),\ldots
,\mathbf{s}_{K}(N)]$, $\mathbf{S}_{U}=[\mathbf{s}_{U}(1),\ldots
,\mathbf{s}_{U}(N)]$, $\mathbf{s}_{K}(t)=[s_{1}(t),\ldots ,s_{P}(
t)]^{T}$ and $\mathbf{s}_{U}(t)=[s_{P+1}(t),\ldots ,s_{D}(
t)]^{T}$.\\
% with $\mathbf{A}\left( \boldsymbol{\theta }_{k}\right)
%=\left[ \mathbf{a}\left( \theta _{1}\right),\ldots
%,\mathbf{a}\left( \theta _{p}\right) \right]$, the known part and
%$\mathbf{A}\left( \boldsymbol{\theta }_{u}\right) =\left[
%\mathbf{a}\left( \theta _{p+1}\right),\ldots,\mathbf{a}\left(
%\theta _{d}\right) \right]$, the unknown one.

%\subsection{Complex unknown gains of each sensor}
%\label{ssec:subhead2}
\textbf{\textit{A2) Complex unknown gains of each sensor}}: The
instrumentation can introduce perturbations such as phase shifts,
in particular due to the difference between sensor gains related
to, e.g., receiver electronics. To correctly specify the model and
avoid inaccurate estimations, calibration needs to be performed.
This can be modeled using the following diagonal calibration
matrix
\begin{equation}
\mathbf{G}=\mathrm{diag}\{\mathbf{g}\}
\end{equation}
where the vector $\mathbf{g} =[g _{1},\ldots ,g _{M}]^{T}$
contains the different unknown complex gains
\cite{vorobyov2005maximum, van2013signal} which
are modeled as DOA independent. %Each gain, for $i{=}1,\ldots ,m$,
%can be written as $\gamma_{i}=g_{i}\exp(j\varphi_{i})$ where
%$g_{i}$ is the strictly positive real amplitude gain and
%$\varphi_{i}$ is the phase that belongs to the interval
%$[0,2\pi[$.
Consequently, the observation matrix (\ref{model2}) can be
rewritten as
\begin{equation}
\label{finalModel}
\mathbf{Y}=\mathbf{G}\mathbf{A}%
(\boldsymbol{\theta}_{K})\mathbf{S}_{K}+\mathbf{G}\mathbf{A}%
(\boldsymbol{\theta}_{U})\mathbf{S}_{U}+\mathbf{N}.
\end{equation}%

%\subsection{Geometry of sensor subarrays in interferometer}
%\label{ssec:subhead3}
\textbf{\textit{A3) Geometry of sensor subarrays}}: In our
scenario, the sensor array is constituted of a set of $L$
subarrays. Due to the large intersubarray distances with respect
to the signal wavelength \cite{zoltowski2000closed,
pesavento2001direction}, the noise is considered statistically
independent between subarrays . Nevertheless, for a given
subarray, sensors being closely spaced, the noise is assumed to be
spatially correlated \cite{vorobyov2005maximum}. Thus, the noise
covariance matrix denoted by $\boldsymbol{\Omega}$ has the
following block-diagonal structure
\begin{equation}
\label{structureOmega} \boldsymbol{\Omega}=
\mathrm{bdiag}\{\boldsymbol{\Omega}_{1},\ldots,
\boldsymbol{\Omega}_{L}\}
\end{equation}
in which $\boldsymbol{\Omega}_{i}$ is a $M_{i}\times M_{i}$ square
matrix where $M_{i}$ is the number of sensors in the $i^{th}$
subarray, such that ${\sum}_{i=1}^{L}M_{i}=M$.\\% The covariance
%matrix $\boldsymbol{\Omega}$ is parameterised by the vector
%$\boldsymbol{\omega }=[\boldsymbol{\omega} _{1}^{T},\ldots
%,\boldsymbol{\omega}_{L}^{T}]^{T}$. Each vector
%$\boldsymbol{\omega}_{i}$ for $i=1,\ldots ,L$ includes the real
%diagonal elements of the noise covariance matrix
%$\boldsymbol{\Omega}_{i}$ and the complex elements whether above
%or below the diagonal of the matrix, due to its
%Hermitian symmetry.\\

\textbf{\textit{Vector of unknown parameters}}: Let us consider a
deterministic/conditional model for the signal sources and
zero-mean complex circular Gaussian noise so that
\begin{equation}
\label{modeldepart}
\mathbf{y}(t)\sim CN\big(\mathbf{G}\mathbf{%
A}(\boldsymbol{\theta})\mathbf{s}(t),\boldsymbol{\Omega}\big).
\end{equation}

Consequently, the vector of unknown parameters is given by
\begin{align}
\nonumber
 \boldsymbol{\eta}=
&%\\ %&
[\boldsymbol{\theta}_{U}^{T}, \mathbf{s}_{U}(1)^{T}, \ldots,
\mathbf{s}_{U}(N)^{T},
\{[\boldsymbol{\Omega}_{1}]_{h_{1},l_{1}}\}_{l_{1}\geq h_{1}},
\ldots,
\\ &
 \{[\boldsymbol{\Omega}_{L}]_{h_{L},l_{L}}\}_{l_{L}\geq
h_{L}}, \mathbf{g}^{T}]^{T}
\end{align}
in which for $i=1,\ldots, L$ and $h_{i}, l_{i} =1,\ldots, M_{i}$,
$\{[\boldsymbol{\Omega}_{i}]_{h_{i},l_{i}}\}_{l_{i}\geq h_{i}}$
represent all the non-zero elements in and above the diagonal of
the noise covariance matrix.

\section{PROPOSED ALGORITHMS}
\label{sec:pagestyle}

In this section, we propose two schemes for joint array
calibration and DOA estimation, based on an iterative ML algorithm
\cite{vorobyov2005maximum, pesavento2001maximum, zhang2015maximum}. The iterative
procedure allows us to obtain a closed-form expression of the
unknown complex gains, the unknown signal sources and the
structured noise covariance matrix. Indeed, the log-likelihood
function is optimized w.r.t. each unknown parameter, while fixing
the others. The different closed-form expressions obtained are
mutually dependent and require an iterative updating procedure
with initialization. For the estimation of the unknown DOAs, an
optimization procedure needs to be performed.

The presence of calibration sources and the iterative procedure
allow us to reduce a $(D+2N
D+{\sum}_{i=1}^{L}M_{i}^{2}+2M)$-dimensional optimization problem
to a $(D-P)$-dimensional optimization problem. The main difference
between the two proposed schemes lies in the estimation of the
calibration matrix as it will be explained in the following.
% We introduce two
%possible methods, based on an iterative procedure. We operate in a
%sequential way \cite{vorobyov2005maximum, pesavento2001maximum},
%in order to obtain an estimation of each unknown parameter with a
%closed-form expression. In the case of $\boldsymbol{\theta}_{U}$,
%an optimisation of a cost function needs to be performed. This
%succession of estimations with closed-form expressions and the
%optimisation problem of $\left(D-P\right)$ dimension enable to
%reduce the computation time. The main difference between the two
%versions presented below, noted the sequential ML based algorithm
%for the first proposed scheme and the sequential modified ML based
%algorithm for the second alternative scheme, lies in the
%estimation of the calibration matrix $\mathbf{G}$.

\subsection{Iterative ML (IML) algorithm for joint array calibration and DOA estimation}
\label{ssec:subhead11}

%We introduce the $m\times 1$ vector $\mathbf{g}$ defined for
%$t{=}1,\ldots ,N$ as:
%\begin{equation}
%\mathbf{g}\left(t\right) =\mathbf{y}(t)-\boldsymbol{\Gamma} \left( \boldsymbol{\gamma }\right)\mathbf{%
%A}(\boldsymbol{\theta })\mathbf{s}(t).
%\end{equation}
%Therefore, we can write the following $m\times N$ matrix:
Let us denote $L(\boldsymbol{\eta})$ the log-likelihood function.
Omitting the constant term, it becomes
\begin{equation}
\label{logv} L(\boldsymbol{\eta})
 ={-N\log}\Big(\det\{
\boldsymbol{\Omega}\}\Big)
 -\mathrm{tr}\left\{\mathbf{V} ^{H}\boldsymbol{\Omega}^{-1}\mathbf{V}
\right\}%+Cst%
%={-N\log }(\det\{ \mathbf{Q}\}) -\sum_{t=1}^{N}
% \mathbf{g}\left(t\right)^{H}\mathbf{Q}^{-1}\mathbf{g}\left(t\right)
%\\ %&
\end{equation}
in which
\begin{equation}
\label{leG}
\mathbf{V} = \mathbf{Y}-\mathbf{G}\mathbf{%
A}(\boldsymbol{\theta})\mathbf{S}.
\end{equation}

\textbf{\textit{1) Estimation of $\boldsymbol{\Omega}$}}: We take
the derivative of $L(\boldsymbol{\eta})$ with respect to the
elements $[\boldsymbol{\Omega}_{i}]_{h_{i},l_{i}}$ for $h_{i},
l_{i}=1,\ldots, M_{i}$ and $i=1,\ldots, L$. During this operation,
all the other unknown parameters remain fixed. We obtain for such
derivation
\begin{align}
\label{derivQ2} \nonumber &\frac{\partial L(\boldsymbol{\eta })
}{\partial[\boldsymbol{\Omega}_{i}]_{h_{i},l_{i}}}= \\ & \nonumber
-\mathrm{tr}\{N\boldsymbol{\Omega}^{-1}\mathbf{e}_{i,h_{i}}\mathbf{e}_{i,l_{i}}^{T}\
-\mathbf{V}^{H}\boldsymbol{\Omega}^{-1}\mathbf{e}_{i,h_{i}}\mathbf{e}_{i,l_{i}}^{T}\boldsymbol{\Omega}^{-1}
\mathbf{V}\}=
\\ &
-N\mathbf{e}_{i,l_{i}}^{T}\boldsymbol{\Omega}^{-1}\mathbf{e}_{i,h_{i}}
+\mathbf{e}_{i,l_{i}}^{T}\boldsymbol{\Omega}^{-1}\mathbf{V}\mathbf{V}^{H}\boldsymbol{\Omega}^{-1}\mathbf{e}_{i,h_{i}}
\end{align}
where
$\left[\mathbf{e}_{i,h_{i}}\right]_{j}=\delta\left(j-h_{i}\right)$
for $j,h_{i}=1,\ldots, M_{i}$ and $i=1,\ldots, L$. Equating
(\ref{derivQ2}) to zero, we obtain the estimations,
$[\hat{\boldsymbol{\Omega}}_{i}]_{h_{i},l_{i}}$, of all the
non-zero elements of $\boldsymbol{\Omega}$. Due to the particular
geometry of sensor subarrays, the exact covariance matrix is
structured as in (\ref{structureOmega}). Consequently, we
introduce $\mathbf{E}=\mathrm{bdiag}\{\mathbf{E}_{M_{1}},\ldots,
\mathbf{E}_{M_{L}}\}$ in order to impose this structure, and the
estimation of $\boldsymbol{\Omega}$ becomes
 %Since we know the
%bloc-diagonal structure of this matrix, we introduce
%$\mathbf{E}=\mathrm{bdiag}\{\mathbf{E}_{M_{1}},\ldots,
%\mathbf{E}_{M_{L}}\}$. Consequently, the closed-form expression
%for the estimation of $\boldsymbol{\Omega}$ in the sequential ML
%based algorithm case is
\begin{equation}
\label{estimQ}
\hat{\boldsymbol{\Omega}}_{\mathrm{IML}}=\frac{1}{N}(\mathbf{V}\mathbf{V}
^{H}) \odot \mathbf{E}.
\end{equation}
One can note that the algorithm can be straightforwardly extended
to the case of other (sparse) colored noise models.\\
%
%For simplicity of writing, we use $\mathbf{A}$, $\mathbf{A}_{K}$,
%$\mathbf{A}_{U}$ and $\mathbf{G}$ instead of
%$\mathbf{A}(\boldsymbol{\theta})$,
%$\mathbf{A}(\boldsymbol{\theta}_{K})$,
%$\mathbf{A}(\boldsymbol{\theta}_{U})$ and
%$\mathbf{G}\left(\mathbf{g}\right)$, respectively.\\

\textbf{\textit{2) Estimation of $\mathbf{G}$}}: We develop the
second part of the r.h.s. of (\ref{logv}) as follows
\begin{align}
\nonumber
 \mathrm{tr}\{  \mathbf{V}
^{H}\boldsymbol{\Omega}^{-1}\mathbf{V}\}=
 \mathrm{tr}\{\mathbf{Y} ^{H}\boldsymbol{\Omega}^{-1}\mathbf{Y}-
\mathbf{Y}
^{H}\boldsymbol{\Omega}^{-1}\mathbf{G}\mathbf{A}(\boldsymbol{\theta})
\mathbf{S}-
\\ %&
\mathbf{S}
^{H}\mathbf{A}(\boldsymbol{\theta})^{H}\mathbf{G}^{H}\boldsymbol{\Omega}^{-1}
\mathbf{Y}+\mathbf{S}
^{H}\mathbf{A}(\boldsymbol{\theta})^{H}\mathbf{G}^{H}\boldsymbol{\Omega}^{-1}\mathbf{G}\mathbf{A}(\boldsymbol{\theta})
\mathbf{S}\}.
\end{align}
Consequently, the derivation of $L(\boldsymbol{\eta }) $ with
respect to the elements $g_{i}$, for $i=1,\ldots ,M$, has the
following form
\begin{align}
\nonumber \label{estimCalib}  \frac{\partial L(\boldsymbol{\eta})
}{\partial g_{i}} = \mathrm{tr}\{\mathbf{Y}
^{H}\boldsymbol{\Omega}^{-1}\mathbf{e}_{i}\mathbf{e}_{i}^{T}\mathbf{A}(\boldsymbol{\theta})\mathbf{S}\
\\ %&
-\mathbf{S}
^{H}\mathbf{A}(\boldsymbol{\theta})^{H}\mathbf{G}^{H}\boldsymbol{\Omega}^{-1}\mathbf{e}_{i}\mathbf{e}_{i}^{T}\mathbf{A}(\boldsymbol{\theta})
\mathbf{S}\} \end{align} where
$\left[\mathbf{e}_{i}\right]_{j}=\delta\left(i-j\right)$, for
$i,j=1,\ldots, M$. Let us denote
$\mathbf{Z}_{1}=\mathbf{A}(\boldsymbol{\theta})\mathbf{S}\mathbf{Y}
^{H}\boldsymbol{\Omega}^{-1}$ and
$\mathbf{Z}_{2}=\mathbf{A}(\boldsymbol{\theta})
\mathbf{S}\mathbf{S} ^{H}\mathbf{A}(\boldsymbol{\theta})^{H}$.
Equating (\ref{estimCalib}) to zero while fixing the other terms
leads us to solve the following linear system of equations
\begin{equation}
\label{equalin} [\mathbf{Z}_{1}]_{i,i}=
[\mathbf{Z}_{2}\mathbf{G}^{H}\boldsymbol{\Omega}^{-1}]_{i,i}, \ \
\ i=1,\ldots ,M.
\end{equation}
Furthermore, let us define the matrix $\mathbf{Z}_{3}$ such that
$[\mathbf{Z}_{3}]_{l,i}=[\mathbf{Z}_{2}]^{\ast}_{l,i}[\boldsymbol{\Omega}^{-1}]^{\ast}_{i,l}$
for $l,i=1,\ldots, M$. In an equivalent way, we can rewrite
(\ref{equalin}) as
\begin{equation}
[\mathbf{Z}_{1}]_{l,l}=\sum_{i=1}^{M}[\mathbf{Z}_{3}]^{\ast}_{l,i}g_{i}^{\ast},
\ \ \ l=1,\ldots ,M.
\end{equation}
Solving this linear system, we obtain for the IML algorithm
\begin{equation}
\label{estimGamma} \hat{\mathbf{g}}_{\mathrm{IML}} =
\mathbf{Z}_{3}^{\dag}\left[ \left[\mathbf{Z}_{1}\right]_{1,1},
\ldots, \left[ \mathbf{Z}_{1}\right]_{M,M}\right]^{H}.
\end{equation}
Consequently,
$\hat{\mathbf{G}}_{\mathrm{IML}}=\mathrm{diag}\{\hat{\mathbf{g}}_{\mathrm{IML}}\}$.\\

\textbf{\textit{3) Estimation of $\mathbf{S}_{U}$}}: Let us denote
$\bar{\mathbf{A}}(\boldsymbol{\theta}_{K})
=\boldsymbol{\Omega}^{-\frac{1}{2}}\mathbf{G}\mathbf{A}(\boldsymbol{\theta}_{K})$,
$\bar{\mathbf{A}}(\boldsymbol{\theta}_{U})=\boldsymbol{\Omega}^{-\frac{1}{2}}\mathbf{G}\mathbf{A}(\boldsymbol{\theta}_{U})$,
 $\tilde{\mathbf{Y}}=\boldsymbol{\Omega}^{-\frac{1}{2}}\mathbf{Y}$,
 $\bar{\mathbf{Y}}=\tilde{\mathbf{Y}}-\bar{\mathbf{A}}(\boldsymbol{\theta}_{K})\mathbf{S}_{K}$ and $\hat{\mathbf{R}}=\frac{1}{N}\bar{\mathbf{Y}}
\bar{\mathbf{Y}}^{H}$.
 The second part of the r.h.s. of (\ref{logv}) can be written as
 \begin{align}
\label{partieint2} \nonumber
 \mathrm{tr}\{\mathbf{V}
 ^{H}\boldsymbol{\Omega}^{-1}\mathbf{V}\}=
%  \\&
%  \nonumber
%\mathrm{tr}\{\big(\boldsymbol{\Omega}^{-\frac{1}{2}}(\mathbf{Y}-\mathbf{G}\mathbf{A}(\boldsymbol{\theta})\mathbf{S})\big)^{H}
% \boldsymbol{\Omega}^{-\frac{1}{2}}(\mathbf{Y}-\mathbf{G}\mathbf{A}(\boldsymbol{\theta})\mathbf{S})\}=
% \\&
 \nonumber
 \mathrm{tr}&\{\bar{\mathbf{Y}}^{H}\bar{\mathbf{Y}}-\bar{\mathbf{Y}}^{H}\bar{\mathbf{A}}(\boldsymbol{\theta}_{U})\mathbf{S}_{U}-
\mathbf{S}_{U}^{H}\bar{\mathbf{A}}(\boldsymbol{\theta}_{U})^{H}\bar{\mathbf{Y}}+
\\&
\mathbf{S}_{U}^{H}\bar{\mathbf{A}}(\boldsymbol{\theta}_{U})^{H}\bar{\mathbf{A}}(\boldsymbol{\theta}_{U})\mathbf{S}_{U}\}.
\end{align}
We take the derivative of $L(\boldsymbol{\eta })$ with respect to
$[\mathbf{S}_{U} ]_{h,l}$, for $h=1,\ldots, (D-P)$ and
$l=1,\ldots, N$ and obtain the estimate in the least squares sense
%\begin{equation}
%\label{pourSu} \frac{\partial L(\boldsymbol{\eta})}{\partial
%[\mathbf{S}_{U}
%]_{h,l}}=\mathrm{tr}\{\bar{\mathbf{Y}}^{H}\bar{\mathbf{A}}(\boldsymbol{\theta}_{U})\mathbf{e}_{h}\mathbf{e}_{l}^{T}\}-
%\mathrm{tr}\{\mathbf{S}_{U}^{H}\bar{\mathbf{A}}(\boldsymbol{\theta}_{U})^{H}\bar{\mathbf{A}}(\boldsymbol{\theta}_{U})\mathbf{e}_{h}\mathbf{e}_{l}^{T}\}.
%\end{equation}
%Equalizing (\ref{pourSu}) to zero leads to the following estimate
%obtained in the least squares sense
\begin{equation}
\label{estimSu}
\hat{\mathbf{S}}_{U}=\Big(\bar{\mathbf{A}}(\boldsymbol{\theta}_{U})^{H}
\bar{\mathbf{A}}(\boldsymbol{\theta}_{U})\Big)
^{-1}\bar{\mathbf{A}}(\boldsymbol{\theta}_{U})^{H}\bar{\mathbf{Y}}.
\end{equation}

\textbf{\textit{4) Estimation of $\boldsymbol{\theta}_{U}$}}:
Plugging (\ref{estimSu}) into (\ref{leG}), we obtain
\begin{equation}
\label{expressG}
\hat{\mathbf{V}}=\boldsymbol{\Omega}^{\frac{1}{2}}\mathbf{P}_{\bar{\mathbf{A}}(\boldsymbol{\theta}_{U})}^{\bot}\bar{\mathbf{Y}}
\end{equation}
in which
$\mathbf{P}_{\bar{\mathbf{A}}(\boldsymbol{\theta}_{U})}^{\bot}=\mathbf{I}-\bar{\mathbf{A}}(\boldsymbol{\theta}_{U})\bar{\mathbf{A}}(\boldsymbol{\theta}_{U})^{\dag}$
is the projector orthogonal to the space spanned by the column
vectors of $\bar{\mathbf{A}}(\boldsymbol{\theta}_{U})$.
%$\mathbf{P}_{\mathbf{\bar{A}}_{u}}^{\bot}=\mathbf{I}-\mathbf{\bar{A}}_{u}(
%\mathbf{\bar{A}}_{u}^{H} \mathbf{\bar{A}}_{u})
%^{-1}\mathbf{\bar{A}}_{u}^{H}$.
Using (\ref{estimQ}) into (\ref{logv}), we can prove that
$\mathrm{tr}\{ \mathbf{V}
^{H}\hat{\boldsymbol{\Omega}}^{-1}\mathbf{V}\}=NM$. Omitting this
constant term and considering the Hermitian symmetry of
$\boldsymbol{\Omega}^{\frac{1}{2}}$ and
$\mathbf{P}_{\bar{\mathbf{A}}(\boldsymbol{\theta}_{U})}^{\bot}$,
one can rewrite
\begin{equation}
L(\boldsymbol{\theta
},\hat{\mathbf{S}}_{U},\hat{\boldsymbol{\Omega}},\mathbf{G})=
{-N\log}\left(\det\left\{\mathbf{Z}\right\}\right)%+\mbox{Cst}
\end{equation}
where we note
$\mathbf{Z}=(\boldsymbol{\Omega}^{\frac{1}{2}}\mathbf{P}_{\bar{\mathbf{A}}(\boldsymbol{\theta}_{U})}^{\bot}\hat{\mathbf{R}}
\mathbf{P}_{\bar{\mathbf{A}}(\boldsymbol{\theta}_{U})}^{\bot}\boldsymbol{\Omega}^{\frac{1}{2}})\odot
\mathbf{E}$. The optimization process is thus
\begin{equation}
\label{estimtheta} \hat{\boldsymbol{\theta}}_{U}=\arg
\min_{\boldsymbol{\theta}_{U}}\Big({\log}\big(\det\left\{
\mathbf{Z}\big\}\right)\Big) .
\end{equation}

\textbf{\textit{Remark}}: To perform the optimization step of the
cost function $F(\boldsymbol{\theta}_{U})={\log }(\det\{
\mathbf{Z}\})$, we use a Newton-type algorithm
\cite{ottersten1993exact}, characterized by a quadratic
convergence. For $l=1,\ldots, (D-P)$, the gradient and the hessian
are given by
\begin{align}
\nonumber \label{gradient} \frac{\partial
F}{[\boldsymbol{\theta}_{U}]_{l}}=\mathrm{tr}\Big\{\mathbf{Z}^{-1}\frac{\partial
\mathbf{Z}}{[\boldsymbol{\theta}_{U}]_{l}}\Big\}\  \ \text{with}
\end{align}
%\begin{align}
%\nonumber
$\frac{\partial \mathbf{Z}}{[\boldsymbol{\theta}_{U}]_{l}}=
\Big(\boldsymbol{\Omega}^{\frac{1}{2}}\Big(\frac{\partial\mathbf{P}_{\bar{\mathbf{A}}(\boldsymbol{\theta}_{U})}^{\bot}}{\partial[\boldsymbol{\theta}_{U}]_{l}}
\hat{\mathbf{R}}\mathbf{P}_{\bar{\mathbf{A}}(\boldsymbol{\theta}_{U})}^{\bot}+
%\\ %&
\mathbf{P}_{\bar{\mathbf{A}}(\boldsymbol{\theta}_{U})}^{\bot}\hat{\mathbf{R}}
\frac{\partial\mathbf{P}_{\bar{\mathbf{A}}(\boldsymbol{\theta}_{U})}^{\bot}}{\partial[\boldsymbol{\theta}_{U}]_{l}}\Big)\boldsymbol{\Omega}^{\frac{1}{2}}\Big)\odot
\mathbf{E}$, and
\begin{equation}
\nonumber \label{hessien} \frac{\partial^{2}
F}{\partial[\boldsymbol{\theta}_{U}]_{l}^{2}}=
\mathrm{tr}\Big\{-\mathbf{Z}^{-1}\frac{\partial
\mathbf{Z}}{[\boldsymbol{\theta}_{u}]_{l}}\mathbf{Z}^{-1}\frac{\partial
\mathbf{Z}}{[\boldsymbol{\theta}_{u}]_{l}}+
\mathbf{Z}^{-1}\frac{\partial^{2}
\mathbf{Z}}{\partial[\boldsymbol{\theta}_{u}]_{l}^{2}}\Big\}\
\text{with}
\end{equation}
%\begin{align}
%\nonumber
$\frac{\partial^{2}
\mathbf{Z}}{\partial[\boldsymbol{\theta}_{U}]_{l}^{2}}=
\Big(\boldsymbol{\Omega}^{\frac{1}{2}}\Big(\frac{\partial^{2}\mathbf{P}_{\bar{\mathbf{A}}(\boldsymbol{\theta}_{U})}^{\bot}}{\partial[\boldsymbol{\theta}_{U}]_{l}^{2}}
\hat{\mathbf{R}}\mathbf{P}_{\bar{\mathbf{A}}(\boldsymbol{\theta}_{U})}^{\bot}+
%\\ %&
\nonumber
2\frac{\partial\mathbf{P}_{\bar{\mathbf{A}}(\boldsymbol{\theta}_{U})}^{\bot}}{\partial[\boldsymbol{\theta}_{U}]_{l}}\hat{\mathbf{R}}
\frac{\partial\mathbf{P}_{\bar{\mathbf{A}}(\boldsymbol{\theta}_{U})}^{\bot}}{\partial[\boldsymbol{\theta}_{U}]_{l}}+ %&
\mathbf{P}_{\bar{\mathbf{A}}(\boldsymbol{\theta}_{U})}^{\bot}\hat{\mathbf{R}}
\frac{\partial^{2}\mathbf{P}_{\bar{\mathbf{A}}(\boldsymbol{\theta}_{U})}^{\bot}}{\partial[\boldsymbol{\theta}_{U}]_{l}^{2}}\Big)
 \boldsymbol{\Omega}^{\frac{1}{2}}\Big)\odot \mathbf{E}$.
%\end{align}
%
%We noted
%$\mathbf{P}^{\bot}=\mathbf{P}_{\bar{\mathbf{A}}(\boldsymbol{\theta}_{U})}^{\bot}$.
%We also introduced the notations
%$\mathbf{P}^{\bot}_{l}=\frac{\partial\mathbf{P}_{\bar{\mathbf{A}}(\boldsymbol{\theta}_{U})}^{\bot}}{\partial[\boldsymbol{\theta}_{U}]_{l}}$
%and
%$\mathbf{P}^{\bot}_{h,l}=\frac{\partial^{2}\mathbf{P}_{\bar{\mathbf{A}}(\boldsymbol{\theta}_{U})}^{\bot}}{\partial
%[\boldsymbol{\theta}_{U}]_{h}\partial[\boldsymbol{\theta}_{U}]_{l}}$,
%\begin{equation}
%\mathbf{P}^{\bot}_{l}=\frac{\partial\mathbf{P}_{\bar{\mathbf{A}}_{u}}^{\bot}}{\partial[\boldsymbol{\theta}_{u}]_{l}}
%\ \ \
%\mathbf{P}^{\bot}_{h,l}=\frac{\partial^{2}\mathbf{P}_{\bar{\mathbf{A}}_{u}}^{\bot}}{\partial
%[\boldsymbol{\theta}_{u}]_{h}\partial[\boldsymbol{\theta}_{u}]_{l}},
%\end{equation}
%The derivations of a projection matrix are precisely presented in
%\cite{ottersten1993exact}.
%\textbf{\textit{Summary of the IML algorithm}}:
\LinesNumberedHidden
\begin{algorithm}
\SetAlgorithmName{IML algorithm}{}{} \caption{}
\SetKwInOut{input}{input} \SetKwInOut{output}{output}
\SetKwInOut{initialize}{initialize}
\input{$\mathbf{Y}$, $\mathbf{E}$, $\mathbf{S}_{K}$,
$\mathbf{A}(\boldsymbol{\theta}_{K})$, $N$, $M$, $L$, $D$, $P$}
\output{estimates of $\boldsymbol{\theta}_{U}$, $\mathbf{S}_{U}$,
$\boldsymbol{\Omega}_{\mathrm{IML}}$ and
$\mathbf{G}_{\mathrm{IML}}$}
\initialize{$\boldsymbol{\Omega}_{\mathrm{IML}}=\mathbf{I}$,
$\mathbf{G}_{\mathrm{IML}}=\mathbf{I}$}  \While
{stop criterion unreached}{ \setcounter{AlgoLine}{0} \ShowLn %%%%% ou stop criterion unreached ou halting criterion false
Estimation of $\boldsymbol{\theta}_{U}$ by (\ref{estimtheta})\\
\ShowLn Estimation of $\mathbf{S}_{U}$ by
(\ref{estimSu})\\
 \ShowLn Estimation of $\boldsymbol{\Omega}_{\mathrm{IML}}$ by
(\ref{estimQ})\\ \ShowLn Estimation of
$\mathbf{G}_{\mathrm{IML}}$ by(\ref{estimGamma})%\\ \ShowLn
%$n=n+1$
\\}
\end{algorithm}

\subsection{Modified iterative ML (MIML) algorithm for joint array calibration and DOA estimation}
\label{ssec:subhead22}

In practical scenario, calibration is performed with respect to
powerful radiating sources. The remaining $(D-P)$ sources, see the
partitioning model in (\ref{partA}) and (\ref{partS}), have a
negligible power in comparison with these calibration sources.
%which is for example the case in the radio astronomy context
%\cite{van2013signal}. This alternative scheme is based on the
%following approximation
%\begin{equation}
%\label{sourcegauss}
%\mathbf{G}\mathbf{A}(\boldsymbol{\theta}_{U})\mathbf{s}_{U}(t)+\mathbf{n}(t)\sim
%CN\left(\mathbf{0},\mathbf{T}\right).
%\end{equation}
Consequently, the distribution of the observations at each
snapshot can be approximated by
\begin{equation}
\label{model} \mathbf{y}(t)\sim CN\Big(
\mathbf{G}\mathbf{A}(\boldsymbol{\theta}_{K})\mathbf{s}_{K}(t),\boldsymbol{\Omega}\Big)
.
\end{equation}
The key idea of this alternative method is to estimate the
calibration parameters and the noise covariance matrix based on
the calibration sources at the first step. Once these parameters
are estimated, the second step consists in estimating the unknown
DOAs and signal sources. For the MIML algorithm, we only present
the results but the methodology is the same as in section
\ref{ssec:subhead11}. Taking into account (\ref{model}) and the
previous calculus performed to obtain (\ref{estimGamma}), we can
estimate $\mathbf{G}$ by solving the following system
\begin{equation}
\label{estimGamma2} \hat{\mathbf{g}}_{\mathrm{MIML}} =
\tilde{\mathbf{Z}}_{3}^{\dag}\left[[ \tilde{\mathbf{Z}}_{1}]
_{1,1}, \ldots, [ \tilde{\mathbf{Z}}_{1}] _{M,M}\right]^{H}
\end{equation}
where
$[\tilde{\mathbf{Z}}_{3}]_{l,i}=[\tilde{\mathbf{Z}}_{2}]^{\ast}_{l,i}[\boldsymbol{\Omega}^{-1}]^{\ast}_{i,l}$
for $l,i=1,\ldots, M$. Here, we have
$\tilde{\mathbf{Z}}_{1}=\mathbf{A}(\boldsymbol{\theta}_{K})\mathbf{S}_{K}\mathbf{Y}
^{H}\boldsymbol{\Omega}^{-1}$ and
$\tilde{\mathbf{Z}}_{2}=\mathbf{A}(\boldsymbol{\theta}_{K})
\mathbf{S}_{K}\mathbf{S}_{K}
^{H}\mathbf{A}(\boldsymbol{\theta}_{K})^{H}$. Consequently,
$\hat{\mathbf{G}}_{\mathrm{MIML}}=\mathrm{diag}\{\hat{\mathbf{g}}_{\mathrm{MIML}}\}$.
%\begin{equation}
%\label{equalin2} [\mathbf{A}_{k}\mathbf{S}_{k}\mathbf{Y}
%^{H}\mathbf{T}^{-1}]_{i,i}= [\mathbf{A}_{k}
%\mathbf{S}_{k}\mathbf{S}_{k}
%^{H}\mathbf{A}^{H}_{k}\mathbf{G}^{H}\mathbf{T}^{-1}]_{i,i}.
%\end{equation}
Following the same methodology to obtain (\ref{estimQ}), the
estimate of $\boldsymbol{\Omega}$ is given by
\begin{equation}
\label{estimT}
\hat{\boldsymbol{\Omega}}_{\mathrm{MIML}}=\frac{1}{N}(\mathbf{V}_{K}\mathbf{V}_{K}
^{H}) \odot \mathbf{E}
\end{equation}
in which $\mathbf{V}_{K} =
\mathbf{Y}-\mathbf{G}\mathbf{A}(\boldsymbol{\theta}_{K})\mathbf{S}_{K}$.

The estimation of the other parameters $\boldsymbol{\theta}_{U}$
and $\mathbf{S}_{U}$ is then performed with the same expressions
as in the first proposed scheme and taking into account the
estimations $\hat{\mathbf{G}}_{\mathrm{MIML}}$ and
$\hat{\boldsymbol{\Omega}}_{\mathrm{MIML}}$ obtained with
(\ref{estimGamma2}) and (\ref{estimT}).

As our simulations will show, the MIML algorithm reaches
convergence faster than the IML algorithm. Furthermore, the latter
requires greater computational complexity, due to the presence of
more estimation steps in the loop.
%
%\subsection{Summary of the two proposed algorithms}
%\label{ssec:subhead33} \LinesNumberedHidden

\begin{algorithm}
\SetAlgorithmName{MIML algorithm}{}{} \caption{}
\SetKwInOut{input}{input} \SetKwInOut{output}{output}
\SetKwInOut{initialize}{initialize}
\input{$\mathbf{Y}$, $\mathbf{E}$, $\mathbf{S}_{K}$,
$\mathbf{A}(\boldsymbol{\theta}_{K})$, $N$, $M$, $L$, $D$, $P$}
\output{estimates of $\boldsymbol{\theta}_{U}$, $\mathbf{S}_{U}$,
$\boldsymbol{\Omega}_{\mathrm{MIML}}$ and
$\mathbf{G}_{\mathrm{MIML}}$}
\initialize{$\boldsymbol{\Omega}_{\mathrm{MIML}}=\mathbf{I}$}
\While
{stop criterion unreached}{ \setcounter{AlgoLine}{0} \ShowLn %%%%% ou stop criterion unreached ou halting criterion false
Estimation of $\mathbf{G}_{\mathrm{MIML}}$ by
(\ref{estimGamma2})\\
\ShowLn Estimation of
$\boldsymbol{\Omega}_{\mathrm{MIML}}$ by (\ref{estimT})%\\
%\ShowLn $n=n+1$
\\}
\setcounter {AlgoLine}{2} \ShowLn Estimation of
$\boldsymbol{\theta}_{U}$ by
(\ref{estimtheta})\\
\ShowLn Estimation of $\mathbf{S}_{U}$  by (\ref{estimSu})\\
\end{algorithm}

%
%\hrulefill
%
%\textbf{Iterative ML based algorithm} %\\
%
%\noindent1. Initialisation:
%$\hat{\boldsymbol{\Omega}}\mathrm{}_{\mathrm{ItML}}=\mathbf{I}$
%and
%    $\hat{\mathbf{G}}_{\mathrm{ItML}}=\mathbf{I}$. \\
%2. Estimation of $\boldsymbol{\theta}_{U}$ by (\ref{estimtheta}).
%\\ 3. Estimation of $\mathbf{S}_{U}$ by (\ref{estimSu}).
%\\
%4. Estimation of $\boldsymbol{\Omega}_{\mathrm{ItML}}$ by
%(\ref{estimQ}).
% \\
%5. Estimation of $\mathbf{G}_{\mathrm{ItML}}$
%by(\ref{estimGamma}).
%\\ 6. Repeat steps 2 to 5 until stop criterion.
%
%\noindent\hrulefill
%
% \textbf{Iterative modified ML based algorithm}\\
%%\textit{\underline{First part}}\\
%1. Initialisation: $\hat{\boldsymbol{\Omega}}_{\mathrm{ItMoML}}=\mathbf{I}$. \\
%2. Estimation of $\mathbf{G}_{\mathrm{ItMoML}}$ by
%(\ref{estimGamma2}).
%\\ 3. Estimation of $\boldsymbol{\Omega}_{\mathrm{ItMoML}}$ by (\ref{estimT}).
%\\ 4. Repeat steps 2 to 3 until stop
%criterion. \\
%%\textit{\underline{Second Part}} \\
%%1.  $\hat{\boldsymbol{\Omega}}=\hat{\mathbf{T}}$.
%%\\
%5. Estimation of $\boldsymbol{\theta}_{U}$ by (\ref{estimtheta}).
%\\
%6. Estimation of $\mathbf{S}_{U}$  by (\ref{estimSu}).
%\\
% 3. Repeat steps 1 to 2 until stop
%criterion.
%
%\noindent\hrulefill\\
%

\section{NUMERICAL SIMULATIONS}
\label{sec:foot}

In the following simulations, we consider two sources, a
calibration source at $\theta_{1}=7^{\circ}$ and an unknown source
at $\theta_{2}=15^{\circ}$, as well as 300 Monte-Carlo and $N=160$
snapshots. The full array  is composed of 3 linear subarrays with
4, 3 and 2 sensors in each. The inter-element spacing is
$\frac{\lambda }{2}$\ in each subarray, 3$\lambda $ and
$\frac{7\lambda }{2}$ between the three successive subarrays. We
consider a noise covariance matrix $\boldsymbol{\Omega}$ with an
identical noise power for each sensor of the same subarray. The
amplitude gains and the phases are generated respectively
uniformly on $\left[ 0,1 \right] $ and $\left[ 0,2\pi \right] $.
% Below is an example of how to insert images. Delete the ``\vspace'' line,
% uncomment the preceding line ``\centerline...'' and replace ``imageX.ps''
% with a suitable PostScript file name.
% -------------------------------------------------------------------------
\begin{figure}[t] %%% ou htb
  \centering
  \centerline{\includegraphics[width=8.5cm, height=6.1cm]{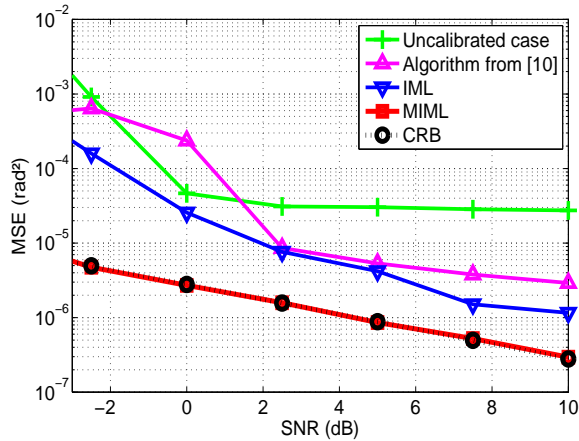}}
%  \vspace{2.0cm}
%  \centering
%  \centerline{\includegraphics[width=4.0cm]{image3}}
%%  \vspace{1.5cm}
%  \centerline{(b) Results 3}\medskip
\caption{Comparison between the IML and the MIML algorithms,
optimizing with Newton.} \label{fig:res1}
\end{figure}
\begin{figure}[t] %%%% ou htb
  \centering
  \centerline{\includegraphics[width=8.5cm, height=6.1cm]{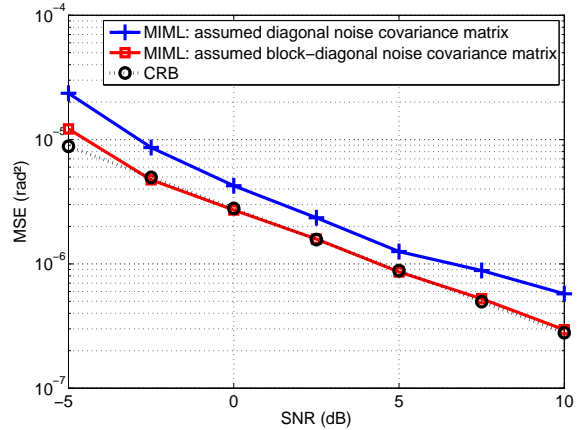}}
%  \vspace{2.0cm}
%  \centering
%  \centerline{\includegraphics[width=4.0cm]{image3}}
%%  \vspace{1.5cm}
%  \centerline{(b) Results 3}\medskip
\caption{Effect of a priori knowledge about the structure of the
noise covariance matrix, for the MIML algorithm.} \label{fig:res2}
\end{figure}
%\begin{minipage}[b]{1.0\linewidth}
%  \centering
%  \centerline{\includegraphics[width=8.5cm]{image1}}
%%  \vspace{2.0cm}
%  \centerline{(a) Result 1}\medskip
%\end{minipage}
%%
%\begin{minipage}[b]{.48\linewidth}
%  \centering
%  \centerline{\includegraphics[width=4.0cm]{image3}}
%%  \vspace{1.5cm}
%  \centerline{(b) Results 3}\medskip
%\end{minipage}
%\hfill
%\begin{minipage}[b]{0.48\linewidth}
%  \centering
%  \centerline{\includegraphics[width=4.0cm]{image4}}
%%  \vspace{1.5cm}
%  \centerline{(c) Result 4}\medskip
%\end{minipage}
%
%\caption{Example of placing a figure with experimental results.}
%\label{fig:res}
%%
%\end{figure}
% To start a new column (but not a new page) and help balance the last-page
% column length use \vfill\pagebreak.
% -------------------------------------------------------------------------
%\vfill \pagebreak
The signal-to-noise ratio (SNR) is denoted by:
\begin{equation}
\mathrm{SNR}=\frac{\sum_{t=1}^{N}\|\mathbf{s}_{U}(t)\|^{2}}{N
M}\sum _{i=1}^{M}\frac{1}{\left[
\boldsymbol{\Omega}\right]_{i,i}}.
\end{equation}

In Fig. \ref{fig:res1}, we plot the mean square error (MSE) vs.
SNR, for both schemes, %We also plot the Cram\'{e}r-Rao bound,
%inspired by Vorobyov \textit{et al} paper
%\cite{vorobyov2005maximum} and Slepian-Bangs formula
%\cite{stoica2005spectral, matveyev1999direction},
 as well as for the uncalibrated case, meaning that the observations are given by
 (\ref{finalModel}) but estimation of matrix $\mathbf{G}$ is not performed in the
 estimation process, it is maintained equal to $\mathbf{I}$. In
this case, we notice the degradation of performances, moreover the
MSE is no longer decreasing from a certain value of the SNR.
%\noindent\textbf{Alternative scheme}\\
%Time (in seconds): 62.782 \\
%Number of iterations: 2 for each part
%
%\noindent\hrulefill
We also plot the MSE of the algorithm proposed in
\cite{vorobyov2005maximum}, for which the presence of calibration
sources is not taken into account. As expected, the presence of a
calibration source enables to achieve better performances,
particularly with the MIML algorithm. The different computation
times for the two exposed methods are 141.473 seconds for the IML
algorithm (4 iterations) and 42.841 seconds for the MIML algorithm
(2 iterations).
%
%\noindent\hrulefill
%
%\noindent\textbf{Iterative ML based algorithm} \ \ \ \ \textbf{Iterative modified ML}\\
%Time (in seconds): 141.473 \ \ \ \ \ \ \ \ \ \ \ 21.902\\
%Number of iterations: 4 \ \ \ \ \ \ \ \ \ \ \ \ \ \ \ \ \ \ 2
%
%\noindent\hrulefill\\

Finally, the Cram\'{e}r-Rao bound (CRB) \cite{vorobyov2005maximum,
stoica2005spectral, matveyev1999direction} was plotted. It is
noticed that the best compromise between computation time and
accuracy of estimation is achieved with the MIML algorithm.
Indeed, we observe that numerically the MSE asymptotically reaches
the CRB. Such performances are due to the estimation of the
calibration matrix $\mathbf{G}$ which is performed separately from
the estimation of $\boldsymbol{\theta}_U$ and mainly depends on
the calibration sources ($\mathbf{A}(\boldsymbol{\theta}_{K})$ and
$\mathbf{S}_{K}$), contrary to the first algorithm
where it depends on the unknown sources as well ($\mathbf{%
A}(\boldsymbol{\theta})$ and $\mathbf{S}$). The IML algorithm
requires more iterations to have better accuracy in the
estimation.
\\ \indent
Finally, Fig. \ref{fig:res2} represents the MSE of the MIML
algorithm for the two following cases: i) taking into account the
true structure of the noise covariance matrix, and ii) assuming
that the noise covariance matrix is diagonal. In the two cases,
the observations are generated using the true noise covariance
matrix which is structured as described by (\ref{structureOmega}).
As expected, such misspecification leads to a higher MSE (case ii)
which shows the importance of taking into account the spatial
correlation due to the array geometry.
% In Fig. \ref{fig:res2}, we plot the MSE vs
%SNR, in the case of the sequential modified ML based algorithm,
%when we are aware of the bloc-diagonal structure of the noise
%covariance matrix $\boldsymbol{\Omega}$ or when we only know it is
%diagonal. This lack of information results in a higher MSE, hence
%the importance of considering such structure, linked to the array
%geometry.

\section{CONCLUSION}
\label{sec:copyright}

In this paper, we proposed two iterative algorithms for joint
calibration and DOA estimation. They are based on the ML
estimation method and are applied in a particular context: some
calibration sources are present, the sensors are characterized by
unknown DOA independent complex gains and the noise covariance
matrix has a block-diagonal structure. The MIML algorithm
outperforms the IML algorithm and numerically attains the CRB for
a low number of iterations. The proposed algorithm is general and
can be adapted, for example, in the context of radio astronomy.
\bibliographystyle{IEEEtran}%\bibliographystyle{IEEEbib}
\bibliography{nab}%\bibliography{strings,refs}

% Generated by IEEEtran.bst, version: 1.13 (2008/09/30)
\begin{thebibliography}{10}
\providecommand{\url}[1]{#1}
\csname url@samestyle\endcsname
\providecommand{\newblock}{\relax}
\providecommand{\bibinfo}[2]{#2}
\providecommand{\BIBentrySTDinterwordspacing}{\spaceskip=0pt\relax}
\providecommand{\BIBentryALTinterwordstretchfactor}{4}
\providecommand{\BIBentryALTinterwordspacing}{\spaceskip=\fontdimen2\font plus
\BIBentryALTinterwordstretchfactor\fontdimen3\font minus
  \fontdimen4\font\relax}
\providecommand{\BIBforeignlanguage}[2]{{%
\expandafter\ifx\csname l@#1\endcsname\relax
\typeout{** WARNING: IEEEtran.bst: No hyphenation pattern has been}%
\typeout{** loaded for the language `#1'. Using the pattern for}%
\typeout{** the default language instead.}%
\else
\language=\csname l@#1\endcsname
\fi
#2}}
\providecommand{\BIBdecl}{\relax}
\BIBdecl

\bibitem{haardt2014subspace}
M.~Haardt, M.~Pesavento, F.~Roemer, and M.~N. El~Korso, ``Subspace methods and
  exploitation of special array structures,'' in \emph{Electronic Reference in
  Signal Processing: Array and Statistical Signal Processing (M. Virberg,
  ed.)}.\hskip 1em plus 0.5em minus 0.4em\relax Academic Press Library in
  Signal Processing, Elsevier Ltd., 2014, vol.~3.

\bibitem{stoica1990maximum}
P.~Stoica and K.~C. Sharman, ``Maximum likelihood methods for
  direction-of-arrival estimation,'' \emph{IEEE Trans. Acoust., Speech, Signal
  Processing}, vol.~38, no.~7, pp. 1132--1143, 1990.

\bibitem{godara1997application}
L.~C. Godara, ``Application of antenna arrays to mobile communications, part
  {II}: {B}eam-forming and direction-of-arrival considerations,''
  \emph{Proceedings of the IEEE}, vol.~85, no.~8, pp. 1195--1245, 1997.

\bibitem{wong1999root}
K.~T. Wong and M.~D. Zoltowski, ``Root-{MUSIC}-based azimuth-elevation
  angle-of-arrival estimation with uniformly spaced but arbitrarily oriented
  velocity hydrophones,'' \emph{IEEE Trans. Signal Processing}, vol.~47,
  no.~12, pp. 3250--3260, 1999.

\bibitem{van2013signal}
A.-J. van~der Veen and S.~J. Wijnholds, ``Signal processing tools for radio
  astronomy,'' in \emph{Handbook of Signal Processing Systems}.\hskip 1em plus
  0.5em minus 0.4em\relax Springer, 2013, pp. 421--463.

\bibitem{ye1995maximum}
H.~Ye and R.~D. DeGroat, ``Maximum likelihood {DOA} estimation and asymptotic
  {C}ram{\'e}r-{R}ao bounds for additive unknown colored noise,'' \emph{IEEE
  Trans. Signal Processing}, vol.~43, no.~4, pp. 938--949, 1995.

\bibitem{friedlander1995direction}
B.~Friedlander and A.~J. Weiss, ``Direction finding using noise covariance
  modeling,'' \emph{IEEE Trans. Signal Processing}, vol.~43, no.~7, pp.
  1557--1567, 1995.

\bibitem{stoica1996maximum}
P.~Stoica, M.~Viberg, K.~M. Wong, and Q.~Wu, ``Maximum-likelihood bearing
  estimation with partly calibrated arrays in spatially correlated noise
  fields,'' \emph{IEEE Trans. Signal Processing}, vol.~44, no.~4, pp. 888--899,
  1996.

\bibitem{li2008maximum}
M.~Li and Y.~Lu, ``Maximum likelihood {DOA} estimation in unknown colored noise
  fields,'' \emph{IEEE Trans. Aerosp. Electron. Syst.}, vol.~44, no.~3, pp.
  1079--1090, 2008.

\bibitem{vorobyov2005maximum}
S.~Vorobyov, A.~B. Gershman, and K.~M. Wong, ``Maximum likelihood
  direction-of-arrival estimation in unknown noise fields using sparse sensor
  arrays,'' \emph{IEEE Trans. Signal Processing}, vol.~53, no.~1, pp. 34--43,
  2005.

\bibitem{stoica1990performance}
P.~Stoica and A.~Nehorai, ``Performance study of conditional and unconditional
  direction-of-arrival estimation,'' \emph{IEEE Trans. Acoust., Speech, Signal
  Processing}, vol.~38, no.~10, pp. 1783--1795, 1990.

\bibitem{KorsoSeq}
M.~N. {El~Korso}, G.~Bouleux, R.~Boyer, and S.~Marcos, ``Sequential estimation
  of the range and the bearing using the zero-forcing {MUSIC} approach,'' in
  \emph{17th European Signal Processing Conf.}, Glasgow, Scotland, 2009.

\bibitem{pesavento2001maximum}
M.~Pesavento and A.~B. Gershman, ``Maximum-likelihood direction-of-arrival
  estimation in the presence of unknown nonuniform noise,'' \emph{IEEE Trans.
  Signal Processing}, vol.~49, no.~7, pp. 1310--1324, 2001.

\bibitem{weiss1990eigenstructure}
A.~J. Weiss and B.~Friedlander, ``Eigenstructure methods for direction finding
  with sensor gain and phase uncertainties,'' \emph{Circuits, Syst., Signal
  Process.}, vol.~9, no.~3, pp. 271--300, 1990.

\bibitem{pierre1991experimental}
J.~Pierre and M.~Kaveh, ``Experimental performance of calibration and
  direction-finding algorithms,'' in \emph{Proc. of {IEEE} Int. Conf. Acoust.,
  Speech, Signal Processing}, Toronto, Ontario, Canada, 1991, pp. 1365--1368.

\bibitem{ng1996sensor}
B.~C. Ng and C.~M.~S. See, ``Sensor-array calibration using a
  maximum-likelihood approach,'' \emph{IEEE Trans. Antennas Propag.}, vol.~44,
  no.~6, pp. 827--835, 1996.

\bibitem{wijnholds2010fish}
S.~J. Wijnholds, ``Fish-eye observing with phased array radio telescopes,''
  Ph.D. dissertation, Technische Universiteit Delft, Delft, The Netherlands,
  2010.

\bibitem{ottersten1993exact}
B.~Ottersten, M.~Viberg, P.~Stoica, and A.~Nehorai, ``Exact and large sample
  maximum likelihood techniques for parameter estimation and detection in array
  processing,'' in \emph{Radar Array Processing}, S.~Haykin, J.~Litva, and
  T.~J. Shepherd, Eds.\hskip 1em plus 0.5em minus 0.4em\relax Berlin:
  Springer-Verlag, 1993, ch.~4, pp. 99--151.

\bibitem{stoica1995concentrated}
P.~Stoica and A.~Nehorai, ``On the concentrated stochastic likelihood function
  in array signal processing,'' \emph{Circuits, Syst., Signal Process.},
  vol.~14, no.~5, pp. 669--674, 1995.

\bibitem{gershman2002stochastic}
A.~Gershman, P.~Stoica, M.~Pesavento, and E.~G. Larsson, ``Stochastic
  {C}ram{\'e}r-{R}ao bound for direction estimation in unknown noise fields,''
  \emph{IEE Proceedings-Radar, Sonar and Navigation}, vol. 149, no.~1, pp.
  2--8, 2002.

\bibitem{chen2008stochastic}
C.~E. Chen, F.~Lorenzelli, R.~E. Hudson, and K.~Yao, ``Stochastic
  maximum-likelihood {DOA} estimation in the presence of unknown nonuniform
  noise,'' \emph{IEEE Trans. Signal Processing}, vol.~56, no.~7, pp.
  3038--3044, 2008.

\bibitem{korsocond}
M.~N. {El~Korso}, R.~Boyer, A.~Renaux, and S.~Marcos, ``Conditional and
  unconditional {Cram\'er-Rao} bounds for near-field source localization,''
  \emph{IEEE Trans. Signal Processing}, vol.~58, no.~5, pp. 2901--2907, 2010.

\bibitem{krim1996two}
H.~Krim and M.~Viberg, ``Two decades of array signal processing research: {T}he
  parametric approach,'' \emph{IEEE Signal Processing Mag.}, vol.~13, no.~4,
  pp. 67--94, 1996.

\bibitem{boyer2008oblique}
R.~Boyer and G.~Bouleux, ``Oblique projections for direction-of-arrival
  estimation with prior knowledge,'' \emph{IEEE Trans. Signal Processing},
  vol.~56, no.~4, pp. 1374--1387, 2008.

\bibitem{bouleux2009optimal}
G.~Bouleux, P.~Stoica, and R.~Boyer, ``An optimal prior-knowledge-based {DOA}
  estimation method,'' in \emph{17th European Signal Processing Conf.},
  Glasgow, United Kingdom, 2009, pp. 869--873.

\bibitem{korso2009fast}
M.~N. El~Korso, R.~Boyer, and S.~Marcos, ``Fast sequential source localization
  using the projected companion matrix approach,'' in \emph{Proceedings of the
  3rd IEEE International Workshop on Computational Advances in Multi-Sensor
  Adaptive Processing}, Aruba, Dutch Antilles, The Netherlands, 2009, pp.
  245--248.

\bibitem{zoltowski2000closed}
M.~D. Zoltowski and K.~T. Wong, ``Closed-form eigenstructure-based direction
  finding using arbitrary but identical subarrays on a sparse uniform
  {C}artesian array grid,'' \emph{IEEE Trans. Signal Processing}, vol.~48,
  no.~8, pp. 2205--2210, 2000.

\bibitem{pesavento2001direction}
M.~Pesavento, A.~B. Gershman, and K.~M. Wong, ``Direction of arrival estimation
  in partly calibrated time-varying sensor arrays,'' in \emph{Proc. of {IEEE}
  Int. Conf. Acoust., Speech, Signal Processing}, Salt Lake City, UT, 2001, pp.
  3005--3008.

\bibitem{zhang2015maximum}
X.~Zhang, M.~N. El~Korso, and M.~Pesavento, ``Maximum likelihood and maximum a
  posteriori {direction-of-arrival} estimation in the presence of {SIRP}
  noise,'' in \emph{Proc. of {IEEE} Int. Conf. Acoust., Speech, Signal
  Processing}, Shanghai, China, 2016.

\bibitem{stoica2005spectral}
P.~Stoica and R.~L. Moses, \emph{Spectral analysis of signals}.\hskip 1em plus
  0.5em minus 0.4em\relax Pearson/Prentice Hall Upper Saddle River, NJ, 2005.

\bibitem{matveyev1999direction}
A.~L. Matveyev, A.~B. Gershman, and J.~F. B{\"o}hme, ``On the direction
  estimation {C}ram{\'e}r-{R}ao bounds in the presence of uncorrelated unknown
  noise,'' \emph{Circuits, Syst., Signal Process.}, vol.~18, no.~5, pp.
  479--487, 1999.

\end{thebibliography}

\end{document}